\documentclass[journal]{IEEEtran}

\usepackage{cite}
\usepackage{graphicx}
\usepackage{amsmath}
\usepackage{url}
\usepackage{xcolor}
\usepackage{cuted}

\newcommand \red[1]{#1}
\newcommand \redtwo[1]{#1}

\begin{document}
\title{Chaos-preserving reduction of the spin-flip model for VCSELs: failure of the adiabatic elimination of the spin-population difference}
\author{Martin~Virte,~\IEEEmembership{Member,~IEEE,}
Francesco~Ferranti,~\IEEEmembership{Senior Member,~IEEE}
\thanks{Manuscript received XXX}
\thanks{This work has been supported by the Research Foundation - Flanders (FWO) through project REFLEX (krediet aan Navorsers, 1530318N) and the post-doctoral fellowship of MV, and by the METHUSALEM program of the Flemish Governments.}%
\thanks{Martin Virte is with the Brussels Photonics (B-PHOT), Dept. of Applied Physics and Photonics, Vrije Universiteit Brussel, Pleinlaan 2, 1050 Brussel, Belgium. He is a post-doctoral Fellow from the Research Foundation - Flanders (FWO). Email: mvirte@b-phot.org}
\thanks{F. Ferranti is with the Microwave Department, Institut Mines-T{\'e}l{\'e}com (IMT) Atlantique, CNRS UMR 6285 Lab-STICC, 29238 Brest CEDEX 3, France. Email: francesco.ferranti@imt-atlantique.fr}}

\maketitle

\begin{abstract}
When studying the dynamics of Vertical-Cavity Surface-Emitting Lasers, and their polarization properties, the spin-flip model appears to be the simplest model qualitatively reproducing all dynamical features that have been observed experimentally. Nonetheless, because of the fast time-scale of the spin-relaxation processes, the specific role and the importance of the spin-population difference - which is one of the specific feature of the spin-flip model - has been continuously questioned. In fact, the debate regarding the possible adiabatic elimination of the spin-population remains fully open. 

In this paper, \red{our goal is to bring new light into this issue by demonstrating} that this variable is essential to preserve the most complex dynamical features predicted by the spin-flip model, such as polarization chaos\red{, and, therefore, needs to be conserved}. To do so, we first perform a detailed analysis, focusing on the chaotic dynamics, to determine the minimal embedding dimension for the spin-flip model. As the latter confirms that a reduction of the model could be envisaged, we then consider the adiabatic elimination of the spin-population difference to highlight and explain its failure to reproduce essential dynamical features obtained in the original model. 
\end{abstract}

\begin{IEEEkeywords}
Semiconductor Laser, VCSELs, chaos, optical chaos, polarization dynamics.
\end{IEEEkeywords}

\section{Introduction}
Due to their intrinsic advantages over standard edge-emitting semiconductor lasers, Vertical-Cavity Surface-Emitting Lasers (VCSELs) are surely making their way towards a widespread use not only in telecom but also for illumination - as in the iPhone X - or gesture recognition and other smart sensing applications. Their main drawback has been identified early on as polarization instabilities \cite{Chang-hasnain1991}: the most striking example being polarization switching (PS) events where the linearly polarized laser light suddenly rotates by $90^\circ$ \cite{Choquette1995, Travagnin1997, Willemsen1999, Willemsen2000, Sondermann2004a}. But a wide range of behaviour have been reported including elliptical polarization \cite{Prati2004}, noise-induced hopping \cite{Nagler2003, Albert2005} and even chaotic dynamics with the so-called polarization chaos \cite{Virte2012, Raddo2017}. Although these instabilities have been first studied to be avoided \cite{Choquette1994, JansenvanDoorn1996a}, it was also remarked that polarization dynamics could be embraced instead: thus leading to the tuning of polarization oscillations for high-speed spin-VCSELs \cite{Pusch2015, Lindemann2016}, polarization switching as optical memory mechanism \cite{Kawaguchi1997, Panajotov1999, Sakaguchi2010} or random bit generation using polarization chaos \cite{Virte2014a}. \\
The polarization features of VCSELs are accurately described using the so-called spin-flip model (SFM) \cite{SanMiguel1995, Martin-Regalado1997}. \red{The main features of the SFM are that} it takes into account: 1) the phase and amplitude anisotropies of the gain medium; 2) two distinct pool\red{s} of carrier depending on their spin, each of them connected to circularly polarized light of a certain handedness. The spin relaxation processes connecting the two carrier pools are then modeled by a simple coefficient commonly called the spin-flip rate. The SFM is, so far, the simplest model qualitatively predicting all the dynamical polarization features experimentally observed in VCSELs. Indeed, other models have been proposed for a given behaviour \cite{Danckaert2002, Nagler2003}, but none - apart from the SFM - reproduced the wide variety of features observed experimentally. On the other hand, the nonlinearities of the SFM hinder the physical interpretation. To get a better understanding and gain further physical insight, different model reductions - based on various approximations - have been proposed \cite{Travagnin1997b, VanExter1998a, Erneux1999}. Experimentally, a rather large value for the spin-flip rate has typically been reported \cite{Al-Seyab2013, Perez2014, Perez2014a}: the spin relaxation rate typically seem\red{s} to be much faster than the other time-scales of the system, in particular the carrier and field decay rates. As a result, adiabatically removing the spin-population difference, i.e. only considering one carrier pool, was often considered to be a reasonable and realistic approximation, especially when focusing on the system steady-states\cite{Travagnin1997b, VanExter1998a, Erneux1999}. On the \red{contrary}, a rather low value of the spin-flip rate seems to be a crucial ingredient to obtain polarization chaos in solitary VCSELs \cite{Virte2017}. Obviously, this observation suggest that, at least for these peculiar experimental cases \cite{Virte2012, Raddo2017}, the spin-population difference could actually play a significant role in the laser dynamics. Hence, we can naturally wonder where is the limit of validity for the approximation allowing the adiabatic elimination of the spin-population difference.\\
In this work, we first aim at confirming that the SFM dimension, i.e. the number of independent variables, could indeed be reduced below the current state of the art. We determine the minimal embedding dimension of the system, i.e. the minimal number of variables required to accurately represent its behaviour, using numerical processing of simulated data based on Principal Component Analysis (PCA) \cite{Jolliffe2002} along with the ''False Neighbors'' technique \cite{Kennel1992, Abarbanel1993}. On top of this important insight, which indicates that a reduction could indeed be considered, we also identify from our analysis data the spin-population difference as a potential candidate for such reduction. Yet, we then show that an adiabatic elimination of the corresponding variable performs, in fact, very poorly at preserving system behavior such as polarization chaos, but also the stability of steady-state solutions. We therefore highlight that the spin-flip model could indeed be further reduced without qualitative loss, but the model-order reduction approach exploiting the adibatic elimination of the spin-population difference is not a suitable solution to accomplish this goal. 

\section{Spin-flip model and parameters}
In its original version, the SFM was a 6-dimension model which has first been introduced in \cite{SanMiguel1995}. When investigating its dynamical properties \cite{Martin-Regalado1997}, it was quickly observed, however, that only the phase difference between the two linear polarization modes played a significant dynamical role. Using phase-amplitude decomposition, it was then proposed to use this phase-difference to get a 5-dimensional model \cite{Erneux1999}. This version of the SFM is the one we exploit here and reads as follows:
\begin{align}
\frac{dR_+}{dt} = &\kappa (N+n-1)R_+ - (\gamma_a cos(\Phi) + \gamma_p sin(\Phi))R_-\\
\frac{dR_-}{dt} = &\kappa(N-n-1)R_- - (\gamma_a cos(\Phi) - \gamma_p sin(\Phi))R_+ \\
\frac{d\Phi}{dt} = &2 \kappa \alpha n - \left(\frac{R_-}{R_+} - \frac{R_+}{R_-}\right) \gamma_p cos(\Phi) \nonumber\\
& + \left(\frac{R_+}{R_-} + \frac{R_+}{R_-}\right)\gamma_a sin(\Phi) \\
\frac{dN}{dt} = &-\gamma\left(-\mu + (N+n)R_+^2 + (N-n) R_-^2\right) \\
\frac{dn}{dt} = &-\gamma_s n - \gamma\left((N+n)R_+^2 -(N-n)R_-^2\right)  \label{eq:carrierpopdifference}
\end{align}
with $R_\pm$ the amplitude of the right and left circular polarizations, $\Phi$ the phase difference between them, $N$ the total carrier population and $n$ the carrier population difference between the two carrier reservoirs for each circular polarization \cite{SanMiguel1995}. Because the carriers in these two reservoirs have different spins, the carrier population difference is often referred to as the spin-population difference. The parameters are as follows: $\kappa$ and $\gamma$ are the field and carrier decay rates respectively, $\gamma_s$ is the spin flip relaxation rate, $\alpha$ is the linewidth enhancement factor, $\gamma_p$ and $\gamma_a$ are the phase and amplitude anisotropies respectively. For simplicity, no misalignment between amplitude and phase anisotropies is considered \cite{Travagnin1997, Virte2014, Virte2015}.\\
Unless stated otherwise, we use the same parameter values as in previous work \cite{Virte2012, Virte2013}: $\kappa = 600 \, ns^{-1}$, $\gamma_s = 100\, ns^{-1}$, $\gamma_a = -0.7 \, ns^{-1}$, $\gamma = 1 \, ns^{-1}$ and $\alpha = 3$. We also consider two distinct cases of chaotic dynamics as discussed in \cite{Virte2013} which are obtained for $\gamma_p = 4 \, ns^{-1}$ (case 1) and $\gamma_p = 25 \, ns^{-1}$ (case 2). The injection current is normalized by the current at threshold, i.e. $\mu=1$ corresponds to the laser threshold when no anisotropies are considered \cite{Martin-Regalado1997}. The normalized injection current $\mu$ is  varied from 1 to 10; the higher limit corresponds\red{, in this theoretical framework,} to 10 times the laser threshold \red{without anisotropies} which is already largely above the current range commonly considered \cite{Raddo2017}. \red{Although more complex dependencies appear to be needed for an accurate estimation of the laser threshold, see e.g. eq. 11 in \cite{Perez2014}, we believe that this higher limit provides a reasonable first order approximation for comparison with experimental observations}. Finally, the linearly polarized (LP) steady-state stable at threshold for the parameters considered here will be identified as X-LP (which is also the low-frequency eigenmode) while the orthogonal LP steady-state will be identified as Y-LP (the high-frequency eigenmode). This {\red{notation}} is identical to the one used in previous work \cite{Virte2012, Virte2013}.\\

\section{Optimal embedding of polarization chaos}
The first part of this work is dedicated to the determination of the minimal number of variables required to accurately describe the polarization chaos dynamics. As such, this will also tell us whether our model could potentially be further reduced and simplified without any loss of generality. A common approach to do so is to use the so-called Principal Component Analysis (PCA) technique\cite{Jolliffe2002}. This technique performs a linear mapping of the data to a lower-dimensional space in such a way that the variance of the data in the low-dimensional representation is maximized. It is a popular technique for dimensionality reduction. The original variables are transformed into new variables, called principal components (PCs), by an orthonormal linear transformation and are ordered by decreasing variance. Using the 5-dimension simulated data generated by the SFM model, we generate a set of 5 linearly independent PCs describing the system trajectory in the variable space. Here, it is however important to note that different pre-processing steps might applied to the data before applying PCA. As any pre-processing step on the data can influence the outcome of the PCA, the user needs to carefully choose them. In this paper, we will use the common centering pre-processing step, i.e. the removal of the mean value of the different variables, and will compare results of the PCA with and without this centering. \\

\begin{figure}
    \centering
    \includegraphics[width=\linewidth]{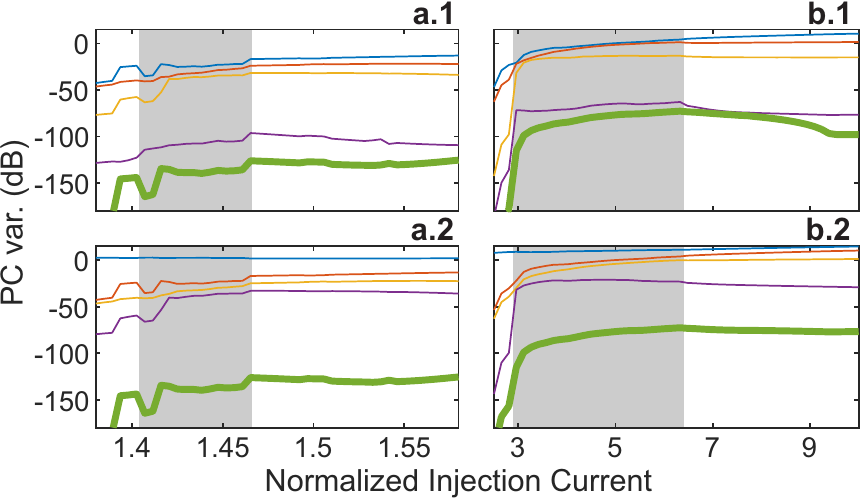}
    \caption{Variance of the 5 PCs of the PCA decomposition for case 1 (a) and 2 (b) for PCA with (a.1, b.1) and without centering (a.2, b.2). In order of decreasing variances, and thus from top to bottom, the PCs are represented in blue, orange, yellow, purple and green. The evolution of the 5th PC is emphasized by a thick line.}
   \label{fig:variance_PC}
\end{figure}

The PCA technique provides a decomposition of the original SFM data matrix $\mathbf{X}$ of size $\# time \ samples$ x $\# states$ (i.e. $\# time \ samples$ x $\# variables$) as
\begin{equation}
\mathbf{P}=\mathbf{X}\mathbf{W}
\end{equation}
\noindent where the matrix $\mathbf{W}$ has dimension $ \# states$ x $\# states$. {\red{The symbol $\#$ is equivalent to "number of".}}  The number of columns of $\mathbf{P}$ and $\mathbf{W}$ can be truncated and a reconstructed $\mathbf{X}_{rec}$ can be computed as
\begin{equation}\label{eq:reconstruction}
\mathbf{X}_{rec}=\mathbf{P}(:,1:r)\mathbf{W}(:,1:r)^{T}=\mathbf{P}_{trunc}\mathbf{W}_{trunc}^{T}
\end{equation}
{\red{where $\mathbf{P}(:,1:r),\mathbf{W}(:,1:r)$ denote the first $r$ columns of the corresponding matrices.}} It is important to highlight that each principal component $\mathbf{P}(:,k), k=1,...,5$ can be expressed as a linear combination as
\begin{equation}\label{eq:linear_comb}
\mathbf{P}(:,k)=\mathbf{X}\mathbf{W}(:,k)
\end{equation}
{\red{where $\mathbf{P}(:,k),\mathbf{W}(:,k)$ denote the $k-th$ column of the corresponding matrices.}} Here, we obtain a set of five linear combinations of variables that are linearly independent of each other in order of decreasing variance. Considering both case 1 ($\gamma_p = 4 \, ns^{-1}$) and case 2 ($\gamma_p = 25 \, ns^{-1}$) for increasing injection currents, we perform PCA independently for each current step. We thus obtain Fig. \ref{fig:variance_PC} that shows the variance associated with each PC after a PCA decomposition with and without centering of the initial data. The regions for which polarization chaos is observed are highlighted by the light gray background. The results are very consistent between case 1 and case 2, but we also see an influence of the centering. In all panels, we clearly see that at least one PC exhibits a variance significantly smaller than the others, more than 50 dB weaker than the variance of the dominant PCs. This therefore suggests that there would indeed be some potential to reduce the number of variables of the model without loss of generality. When the centering pre-processing is used, two PCs - instead of only one - show smaller variance than the dominant PCs. Interpreting this result as a possible reduction of two orders, i.e. possible reduction to a 3D model, might be an overstatement, but these results strongly suggest that at least one variable could be removed. \\ 

When dealing with chaotic dynamics, it is obvious that any small changes of the system should be considered with the highest level of caution as these could have a dramatic impact on the system behaviour. To verify the suitability of an embedding for a chaotic systems, a common test is the so-called ''False Neighbors'' test \cite{Kennel1992, Abarbanel1993}. If the embedding dimension is too small, the chaotic attractor will necessarily be folded, i.e. some parts of the chaotic trajectory will artificially become close and thus create the so-called ''False Neighbors''. To detect such folding, we use the following approach. After applying PCA, we reconstruct the 5-D states trajectories using only a subset of the calculated PCs using (\ref{eq:reconstruction}). We then compare the distance between the data point and its closest neighbor in the initial and reconstructed 5-dimension state trajectories space. From these data, we derive two figures of merit:
\begin{itemize}
    \item the Mean Distance Growth (MDG): the average growth for all selected closest neighbors for all data points considered. Obviously, a large MDG $>> 1$ indicates a probable folding while a MDG very close to 1 will tend to confirm that the embedding dimension is large enough to contain the chaotic trajectory.
    \item the amount of ``False Neighbors'' (FN): the number of closest neighbors - for all data points - for which the distance growth exceed a given threshold $R_{tol}$. The amount of FN of course depends on the threshold value, but, as discussed in \cite{Kennel1992}, the approach is quite robust against the choice of threshold. Here, we show results for $R_{tol} = 20$, but verified that the same conclusions hold for threshold values from 2 to 30.  
\end{itemize}

\begin{figure}
    \centering
    \includegraphics[width=\linewidth]{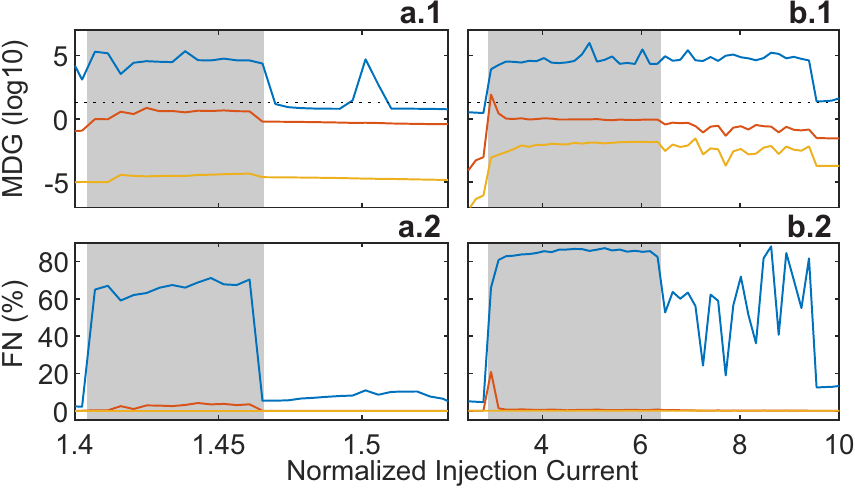}
    \caption{Mean Distance Growth and False Neighbor test results with reconstruction based on PCA \textbf{without} centering for case 1 (a) and case 2 (b). For each panel, we show the embedding of 2 (blue, top curve), 3 (red, middle curve) and 4 (yellow, bottom curve) PCs for an increasing injection current. (a.1, b.1) Mean Distance Growth minus unity shown in logarithmic scale. $0$ corresponds to a doubling of the nearest neighbors distance while $-\infty$ correspond to no distance increase. The dotted horizontal lines indicate the threshold value used to detect false neighbors. (a.2, b.2) Ratio of False Neighbors detected over the whole time-series considered.}
    \label{fig:FNtest_POD}
\end{figure}

\begin{figure}
    \centering
    \includegraphics[width=\linewidth]{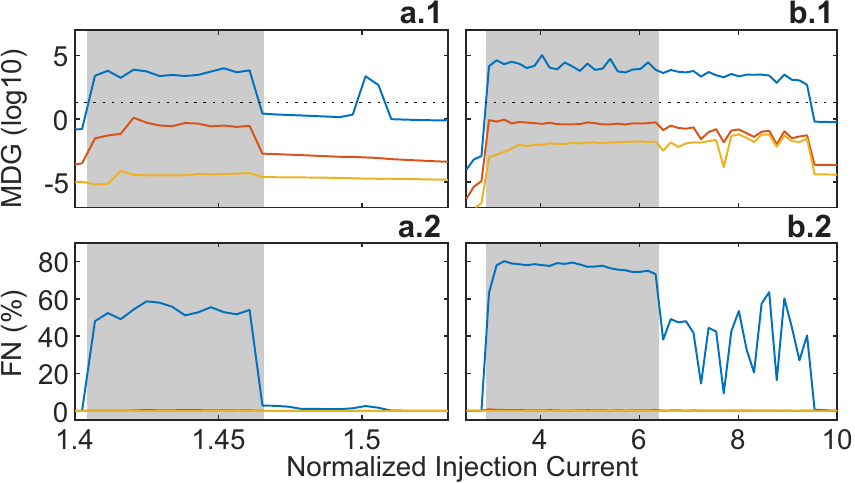}
    \caption{Mean Distance Growth and False Neighbor test results with reconstruction based on PCA \textbf{with} centering. The rest of the caption is identical to the one of Fig. \ref{fig:FNtest_POD}}
    \label{fig:FNtest_PCA}
\end{figure}

\begin{figure}
    \centering
    \includegraphics[width=\linewidth]{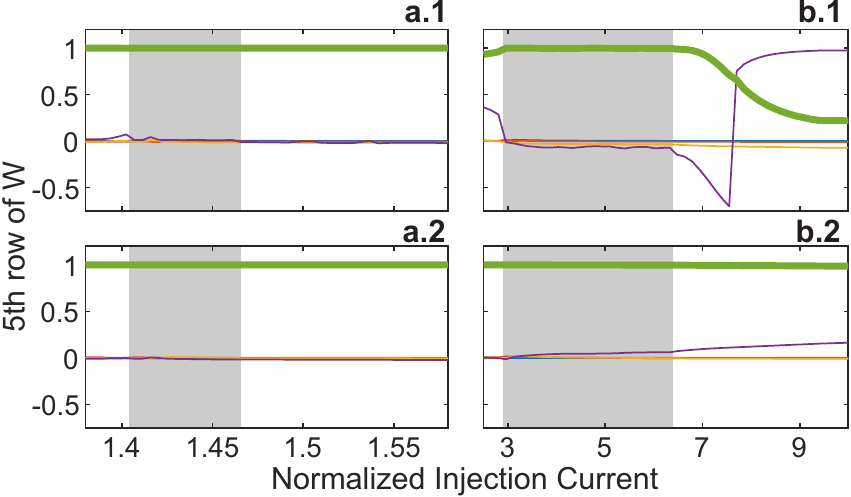}
    \caption{Evolution of the contribution of the spin-population difference $n$ to each of the 5 PCs - i.e. the fifth row of the $\mathbf{W}$ matrix - for case 1 (a) and 2 (b). In order of decreasing variances, the PCs are represented in blue, orange, yellow, purple and green. The 5th PC is emphasized by a thick line. The top panels (a.1) and (b.1) show the evolution for the PCA with centering, while (a.2) and (b.2) give their counterpart obtained without centering.}
    \label{fig:W_5th_row}
\end{figure}

Here, we consider 10000 data point per time-series with a time-step of 25 ps. We take the single closest neighbor for each data point \cite{Kennel1992} and impose a minimal time separation of 1000 ps between them to avoid correlation in time. The evolution of these two figures of merit for increasing currents and for different embedding dimensions are shown in Figs. \ref{fig:FNtest_POD}-\ref{fig:FNtest_PCA} without and with centering, respectively. Considering both data sets, it is clear that considering only the first 2 PCs is largely insufficient as the amount of false neighbors is quite large: for both cases, inside the chaotic region, the closest neighbor after embedding is most of the time a "false" neighbor. Adding the third PC induces a significant improvements, and only a few false neighbors are detected - between 1 to 4 \% without centering (Fig. \ref{fig:FNtest_POD}) and less than 0.5 \% with centering (Fig. \ref{fig:FNtest_PCA}) - and only inside the chaotic region. But it is only when considering the first 4 PCs that the amount of detected False Neighbors effectively reach and remain at 0 for all conditions. From a system analysis viewpoint, the imperfect embedding using only 3 PCs could be considered for a first approximation even though it would miss some details of the dynamics. On the other hand, from a nonlinear dynamics viewpoint, a 4 dimension embedding using the first 4 PCs seems to be sufficient to contain the whole chaotic attractor without inducing any detectable folding. Indeed, in this case absolutely no false neighbors is detected and low value of the mean distance growth is systematically observed. In essence, these {\red{results}} confirm that the 5th PC, i.e. the PC with the smallest variance, does not provide essential information over the system behaviour or chaotic features. Therefore, we can conclude that these results confirm that the spin-flip model could be reduced from 5 to 4 variables while preserving all complex features including chaos.\\ 
Considering this perspective, we can further analyze the different outputs of the PCA to identify the variables that could effectively be dismissed. Such information can be provided by the $\mathbf{W}$ matrix which describe{\red{s}} the linear combination of the original variables leading to the PC description. To put it differently, the $\mathbf{W}$ matrix shows the contribution of each variable to the different PCs. Most variables are actually contributing to different PCs in a rather complex way. In Fig. \ref{fig:W_5th_row}, we show the contribution of the spin-population difference $n$ to each of the PCs. Thus, it is rather striking that the spin-population difference is, in fact, almost only contributing to the 5th PC, with the exception of Fig. \ref{fig:W_5th_row}(b.1) that we discuss below. It is however important to understand that this does not mean that the 5th PC is almost perfectly equal to the $n$ variable, since the other SFM variables can also have an impact on the 5th PC in the linear combination formula (\ref{eq:linear_comb}). The result shown in Fig. \ref{fig:W_5th_row} suggests that the spin-population difference $n$ could be a good candidate to be considered for the model reduction as we will do in the next section using an adiabatic elimination. The failure of this adiabatic elimination approach, which we will discuss in what follows, should however not lead to the conclusion that the SFM cannot be reduced to 4 variables despite the outcome of the PCA and "False Neighbors" analyses, but only to the fact that another approach needs to be considered to obtain a suitable model-order reduction. 
Finally, in Fig. \ref{fig:W_5th_row}(b.1), i.e. for PCA with centering of case 2: in this case, the spin-population difference also appears to contribute to the 4th PC shown in purple. Yet, this result needs to be put in perspective with the variance of this 4th PC as shown in Fig. \ref{fig:variance_PC}(b.1): there, we see that the 4th and 5th PC both have very low variance, more than 50 dB below the three other PCs. Hence it does not invalidate the proposed interpretation as the spin-population difference $n$ never contributes to one of the dominant PC. Nevertheless, this might eventually be seen as a hint of the failure of the adiabatic elimination reduction, but this interpretation would be far too speculative at this stage. 

\section{Adiabatic elimination of the carrier population difference}
As discussed above, the dynamical analysis that we performed suggests that the spin-population difference $n$ could be a good candidate for the reduction of the spin-flip model. Since a similar conclusion can be reached from physical considerations \cite{Travagnin1997b, VanExter1998a}, this approach obviously need to be further investigated. \\

\begin{figure}
    \centering
    \includegraphics[width=\linewidth]{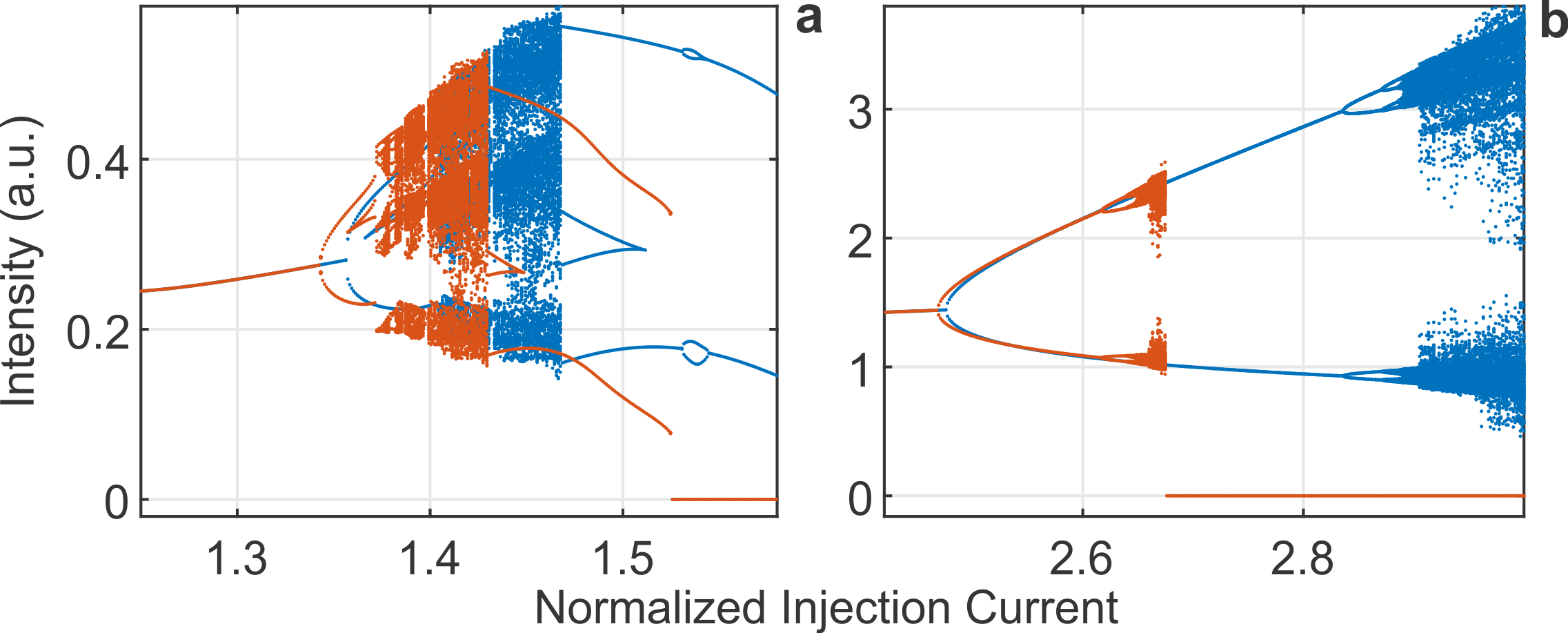}
    \caption{Comparison of bifurcation diagrams between the full SFM (blue) and the reduced model (orange) after adiabatic elimination of the carrier population difference for case 1 (a) and 2 (b), respectively. The diagram shows the maxima of the intensity of X-LP, i.e. the linear polarization stable at threshold.}
    \label{fig:diagcompare}
\end{figure}

Based on the rate equations describing the spin-flip model, and similarly to what has been done in previous works \cite{Travagnin1997b, VanExter1998a}, we can eliminate the carrier population difference adiabatically. Starting with (\ref{eq:carrierpopdifference}) and considering that the spin-flip rate $\gamma_s$ is large enough for $n$ to reach a steady-state much faster than the other variables, we directly obtain that:
\begin{equation}
    n = \frac{N (R_-^2 - R_+^2)}{\gamma_s/\gamma + R_+^2 + R_-^2} \label{eq:expr_n_elimination}
\end{equation}
The differences with the expression found in \cite{VanExter1998a} are only due to the particular approximation made therein and that we do not apply here. We can then easily confirm that this expression of the carrier population difference is an accurate approximation by comparing it to the simulated values. By doing so, we obtain that the error is typically two-orders of magnitude smaller than the value taken by the carrier population difference variable. As such, although such small error will have an obvious impact on the simulated chaotic time-series, we would expect at this point a negligible qualitative impact on the laser behaviour. On the contrary, when computing bifurcation diagrams - as shown in Fig. \ref{fig:diagcompare} - using this reduced model, we can only observe rather dramatic qualitative changes. In particular, in case 2, the initially wide chaotic region is shifted and shrunk into a narrow range of injection current. Thus, for currents above 2.7, the laser only exhibits a stationary behaviour corresponding to Y-LP, i.e. the linearly polarized steady-state orthogonal to the steady-state stable at threshold X-LP. This is particularly puzzling considering that, in case 2, the Y-LP steady-state is supposed to be unstable up to extremely high values of current ($\mu>50$) \cite{Virte2013}.\\

\begin{figure}[t]
    \centering
    \includegraphics[width=0.95\linewidth]{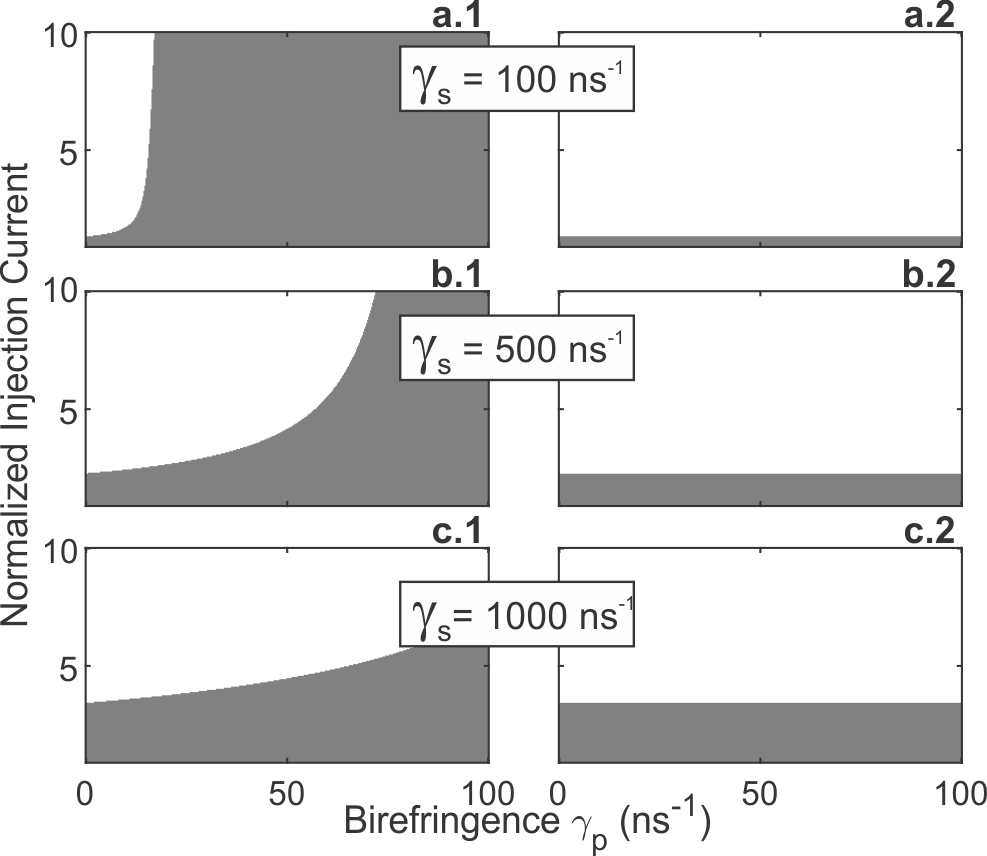}
    \caption{Stability map of the Y-LP steady-state as a function of the birefringence and the injection current for the full SFM (1, left) and the reduced model (2, right) for three distinct values of spin-flip rate: $100 \, ns^{-1}$ (a), $500 \, ns^{-1}$ (b) and $1000 \, ns^{-1}$ (c). The white (grey) regions indicate where the Y-LP steady state is stable (unstable).}
    \label{fig:stabilitymap}
\end{figure}

To clarify this issue, we have re-investigated the stability of both the X-LP and Y-LP steady-state for the full SFM and for the reduced model. To compute it, we linearize the rate equations around the steady-states and numerically extract the eigenvalues of the linear system. By adiabatically eliminating the carrier population difference, we obviously remove one equation. But using the expression of $n$ given in (\ref{eq:expr_n_elimination}) also creates several additional dependencies that are nonexistent in the full model. While no impact of the adiabatic elimination is observed on the stability of the X-LP state 
\red{(not shown), we observe that the stability of the Y-LP state is not well-preserved. As displayed in Fig. \ref{fig:stabilitymap}, the lower boundary of the stability region with respect to the injection current appears to be independent of the birefringence for the reduced model while a clear dependence can be observed in the full-model. As could be expected, this discrepancy is reduced as the value of the spin-flip rate is increased, but this agreement is still imperfect and restricted to low values of the birefringence $\gamma_p$.} \\
\red{We now investigate analytically the stability of the Y-LP steady-state. As described in the appendix, the linearized system has two pairs of eigenvalues $\lambda_1$ and $\lambda_2$ of the form $A\pm\sqrt{B}$. Because $\lambda_2$ always corresponds to a pair of complex conjugated eigenvalues with a negative real part for the typical range of parameter values that we consider here, we will focus on $\lambda_1$ which effectively limits the stability of Y-LP. The latter can be expressed as follows: 
\begin{equation}
    \lambda_1 = \left( - 2\gamma_a - \chi \right) 
    \pm \sqrt{\chi^2 - 4\alpha\gamma_p \chi - 4 \gamma_p^2 }
    \label{eq:lambda1}
\end{equation}
with the variable $\chi$ being defined as:  
\begin{equation}
    \chi = \frac{\gamma \kappa}{\gamma_s} \left( \mu - 1 + \frac{\gamma_a}{\kappa} \right)
\end{equation}
In practice, $\chi$ corresponds to the injection current, normalized by the Y-LP steady-state threshold $\mu_{Y-LP,th} = 1 - \gamma_a/\kappa$ - meaning that $\chi$ is necessarily positive - and with a peculiar scaling.
Inside the square root term in (\ref{eq:lambda1}), we have a convex $2^{nd}$ order polynomial equation in $\chi$. The roots of the equation are $2\gamma_p (\alpha - \sqrt{\alpha^2+1})$ and $2\gamma_p (\alpha + \sqrt{\alpha^2+1})$, which are respectively negative and positive for $\gamma_p > 0$. In this case, the second term represents a crucial threshold with respect to $\chi$: above it, we have real eigenvalues, while, below it, we have a pair of complex conjugated eigenvalues. Thus, it is important to remark that, with the set of parameters considered in this work, this threshold is increasing along with $\gamma_p$ and is already above $\mu=5$ for $\gamma_p = 2 \, ns^{-1}$. As a result, for most cases, the latter will apply, and the stability is then defined by the first term of ( \ref{eq:lambda1}), i.e. the real part of the eigenvalue pair. From its expression, we immediately get that the real part is negative if:
\begin{equation}
\mu > 1 - \frac{\gamma_a}{\kappa} - 2\frac{\gamma_a \gamma_s}{\kappa \gamma} \label{eq:YLP_lower_bound}
\end{equation}
This equation is almost the same as eq. 55 of \cite{Erneux1999}, except for the 2nd term $\gamma_a/\kappa$ which is however expected to have a negligible impact as we typically have $\gamma_a << \kappa$. This is clearly the stability limit for the Y-LP state visible in Fig. \ref{fig:stabilitymap} which correspond to a well-known Hopf bifurcation \cite{Martin-Regalado1997, Erneux1999}. \\
When we have two real eigenvalues, since the square root term is necessarily positive, the largest eigenvalue will necessarily be the one for which the two terms are added. In this case, it can only be negative if the following condition is fulfilled: 
\begin{equation}
\sqrt{\chi^2 - 4\alpha\gamma_p \chi - 4 \gamma_p^2} < 2\gamma_a + \chi
\end{equation}
To have a real eigenvalue, the left hand side must be positive. This implies that $2\gamma_a + \chi > 0$ - which corresponds to (\ref{eq:YLP_lower_bound}) - is a necessary condition for the previous inequality to be verified. If both sides are positive, the previous equation is equivalent to:
\begin{equation}
    0< (\gamma_a^2 + \gamma_p^2) + (\gamma_a + \alpha \gamma_p)\chi
\end{equation}
If $(\gamma_a + \alpha \gamma_p)>0$, this inequality is obviously always verified. For $\gamma_a<0$ and $\gamma_p >0$, this comes down to a threshold on the birefringence: $\alpha \gamma_p > |\gamma_a|$, which, with our typical set of parameters, corresponds to values of the birefringence above approximately $0.23$. On the other hand, when $\gamma_a + \alpha \gamma_p <0$ the Y-LP steady-state appears to be only stable when:
\begin{align}
    \chi &< - \frac{\gamma_a^2 + \gamma_p ^2}{\alpha \gamma_p + \gamma_a}\\
    \text{i.e.} \quad \mu &< 1 - \frac{\gamma_a}{\kappa} - \frac{\gamma_s (\gamma_a^2 + \gamma_p^2)}{\gamma\kappa(\alpha \gamma_p + \gamma_a)} 
\end{align}
Apart from the extra $\gamma_a/\kappa$ term which is expected to be negligible, this expression is similar to eq. 53 found in \cite{Erneux1999} but for the X-LP steady state and with a change of sign. This difference is easily explained by the fact that X-LP and Y-LP are interchangeable by simultaneously changing the sign of $\gamma_a$ and $\gamma_p$. This leads to the same change of sign, as shown in fig. 1 (a) and (b) of \cite{Erneux1999}, where the same curve is obtained for $\mu_{xs}$ and $\mu_{ys}$. Using the same approach as for Fig. \ref{fig:stabilitymap}, we could also confirm that an excellent agreement is obtained between the reduced and complete model for $\gamma_p < \gamma_a/\alpha$.\\
To conclude, for the case of $\gamma_a<0$ and $\gamma_p >0$, the stability of the Y-LP steady-state is largely impacted by the adiabatic elimination of the spin-population inversion. While for sufficiently small values of the birefringence $\gamma_p$, an acceptable agreement between the reduced and complete model can be obtained, significant discrepancies arise when $\gamma_p$ is increased. In the reduced model as soon as $\gamma_p$ is larger than $\gamma_a/\alpha$, the Y-LP state is always stable when the condition of eq. \ref{eq:YLP_lower_bound} is met. This means that, intrinsically, the stability of Y-LP becomes independent of $\gamma_p$. Although this reduction could be a sufficient but limited approximation for very large spin-flip rates $\gamma_s > 1000 \, ns^{-1}$ and small birefringence values $\gamma_p \approx 5 \, ns^{-1}$, we showed in our previous work \cite{Virte2017} that these are, by far, conditions not suitable to generate polarization chaos dynamics. 
}

\section{Discussion}
Looking at all in-depth investigations of VCSELs dynamics, it is undeniable that the spin-flip rate is an essential parameter controlling the dynamical behaviour of VCSELs. Yet, the spin-flip rate only plays a role of the laser dynamics through its influence on the spin population difference. Often, it is assumed - also considering the typically large values of the spin-flip rate reported experimentally - that the adiabatic elimination of the spin-population difference could be a suitable approximation and would retain all essential non-linearities in the SFM model. Yet, in this work, we highlight that the impact of this adiabatic elimination on the polarization chaos and stability of the system steady-states is far too large to be neglected, even for large values of the spin-flip rate and, especially, for birefringence values in the order of tens of $ns^{-1}$ and above \cite{Pusch2015, Lindemann2016}. However, our analysis using the PCA and "False Neighbors" techniques confirm that there is potential for model order reduction, but it needs to be performed using methods that, unlike the adiabatic elimination of the spin-population, would preserve all key dynamical features of the model. However, is is an aspect that we will leave for future work. {\red{Model order reduction is a vast field \cite{Antoulas05,schilders08,quarteroni14,Benner17}. Depending on the complexity of the system representation (e.g., linear time-invariant, time-variant, non linear, etc.) to be reduced, the model order reduction approach has an increasing complexity especially if system properties need to be preserved in the reduced model. Model order reduction schemes very often use specific mathematical transformation to transform the original state vector space into a reduced space, while keeping accuracy and properties with respect to the original model. The property-preserving model order reduction is much more complex than just an accuracy-preserving model order reduction. Already in the case of linear time-invariant systems, model order reduction techniques can provide accurate reduced order models of the transfer function behavior in a frequency bandwidth of interest, but they can fail in preserving properties such as stability and passivity. For nonlinear systems, it is obviously even more complex. Therefore, a dedicated investigation of such techniques that can perform model order reduction of the original SFM equations are beyond the scope of this contribution and deserves a dedicated effort in a separated work.}} \\
Finally, we would like to discuss why the work presented here leads to conclusions that significantly differ from previous reports, in particular \cite{Travagnin1997b, VanExter1998a}. We believe that this can be mostly attributed to the different focus of our analysis: while the authors of \cite{Travagnin1997b, VanExter1998a} were working to get a better physical understanding of the stability boundaries of the linearly polarized steady-states, we look here at the modelling accuracy of the dynamical features. Similarly, while strong approximations are taken in \cite{Travagnin1997b, VanExter1998a}, we tried to keep a very general viewpoint, thus did not use similar approximation. In the end, we are convinced that these two aspects are of course complementary. However, as highlighted in this work, when considering dynamical behaviours, the spin-population difference appears as an essential piece of the puzzle that must be taken into account to its full extent to ensure qualitative relevance of the modelling. 

\red{
\appendices
\section{Stability of the Y-LP steady-state in the adiabatically reduced spin-flip model}
The Y-LP steady-state, is defined as follows: 
\begin{equation}
N = 1 - \frac{\gamma_a}{\kappa} \quad,
\phi = \pi \quad,
R = \sqrt{\frac{\mu-N}{2N}}
\end{equation}
For simplicity, we will approximate the expression of the adiabatically eliminated population difference by considering that the denominator can be directly expressed as $\gamma_s /\gamma$. Since this term is typically of the order of 100 to 1000 while the amplitude terms are expected to be of the order of 1, the approximation is small for the practical cases considered here. After linearization of the equations, we obtain the following system whose eigenvalues will determine the stability of the steady-state.

\begin{strip}
\begin{equation}
    \begin{bmatrix}
    \Dot{\delta R_+} \\
    \Dot{\delta R_-} \\
    \Dot{\delta \phi} \\
    \Dot{\delta N}
    \end{bmatrix}
   =
    \begin{bmatrix}
    \kappa(N-1) - 2 \kappa\gamma N R^2/\gamma_s  
    & \gamma_a + 2\kappa\gamma N R^2/\gamma_s   
    & \gamma_p R 
    & \kappa \gamma R\\
    
     \gamma_a + 2\kappa\gamma N R^2/\gamma_s 
    & \kappa(N-1) - 2 \kappa\gamma N R^2/\gamma_s
    & -\gamma_p R
    & \kappa \gamma R\\
    
    -4\kappa\gamma\alpha N R/\gamma_s - 2\gamma_p/R
    & 4\kappa\gamma\alpha N R/\gamma_s + 2\gamma_p/R
    & -2\gamma_a
    & 0 \\
    
     -2\gamma N R
    & - 2\gamma N R
    & 0
    & -\gamma(1+2R^2)\\
    \end{bmatrix}
    \begin{bmatrix}
    {\delta R_+} \\
    {\delta R_-} \\
    {\delta \phi} \\
    {\delta N}
    \end{bmatrix}
\end{equation}
\end{strip}
Using a symbolic computation software, we obtain two pairs of eigenvalues of the form $\lambda_1 = A1 \pm \sqrt{B1}$ and $\lambda_2 = A2 \pm \sqrt{B2}$. Using the following change of variable 
\begin{equation}
    \chi = \frac{\gamma \kappa}{\gamma_s} \left( \mu - 1 + \frac{\gamma_a}{\kappa} \right)
\end{equation}
the first eigenvalue pair can be efficiently expressed as follows: 
\begin{equation}
    \lambda_1 = \left( - 2\gamma_a - \chi \right) 
    \pm \sqrt{\chi^2 - 4\alpha\gamma_p \chi - 4 \gamma_p^2 }
\end{equation}
The second pair of eigenvalue can be expressed as: 
\begin{equation}
    \lambda_2 = \frac{\gamma \kappa \mu}{2(\gamma_a - \kappa)} \pm \sqrt{\frac{\gamma^2\kappa^2\mu^2}{4(\gamma_a - \kappa)^2} - 2\gamma\kappa\mu +  2\gamma\kappa(\frac{\gamma_a}{\kappa}-1)} 
\end{equation}
As mentioned in the core of the article, $\lambda_1$ is the interesting pair of eigenvalue and is therefore further described therein. The impact of $\lambda_2$ is far more limited as will be described here.\\
Since we typically have $|\gamma_a| << \kappa$ with $\kappa$ being positive by design, the first term will always be negative for the considered parameter range. \redtwo{If the square root is imaginary - which is typically the case for the standard parameter range considered here -, we obtain two complex conjugate eigenvalues with negative real part. On the other hand, if it is real, its absolute value is smaller than the absolute value of the first term, and we therefore have two negative real eigenvalues. Either way, the system will not be destabilize by this second pair of eigenvalue.}}


\begin{IEEEbiographynophoto}{Martin Virte}
received the master in engineering with a major in Photonics from the French “Grande Ecole” Sup\'{e}lec (now CentraleSup\'{e}lec, Universit\'{e} de Paris-Saclay, France) and the M.Sc. in Physics from Sup\'{e}lec and the Universit\'{e} de Lorraine both in 2011. He simultaneously obtained the PhD degree in Photonics from Sup\'{e}lec and in engineering from the Vrije Universiteit Brussels, Brussels, Belgium in 2014 in the frame of a joint PhD. In 2015, he received a three-year post-doctoral fellowship from the Research Foundation Flanders (FWO) to pursue his research at the Vrije Universiteit Brussel, Belgium. Since 2018 he is research professor with the Brussels Photonics Team (B-PHOT) of the Vrije Universiteit Brussel. \\
Martin authored and co-authored more than 16 papers in international refereed journals and several contributions in national and international conferences. Martin received the Graduate Student Fellowship Award from the IEEE Photonics Society in 2014 in recognition of his PhD work. His research is currently focusing on laser dynamics, more competition, optical chaos and their applications in sensing and communication. In addition, he acts as a reviewer for several international journals including journal of lightwave technology, chaos, applied physics letters and others. He is a member of SPIE, IEEE (IEEE Photonics Society) and OSA. 
\end{IEEEbiographynophoto}

\begin{IEEEbiographynophoto}{Francesco Ferranti}
(M'10-SM'17) received the Ph.D. degree in electrical engineering from Ghent
University, Ghent, Belgium, in 2011. He is currently an Associate Professor at the Microwave Department at Institut Mines-T{\'e}l{\'e}com (IMT), Brest, France. He has been awarded the Anile-ECMI Prize for
Mathematics in Industry 2012 and the Electromagnetic Compatibility Society President's Memorial
Award 2012.

He has authored and co-authored 56 papers in international peer-reviewed journals, 52 papers in international peer-reviewed conferences and 2 books chapters. He has given invited lectures and
chaired several sessions at international conferences. He serves as a regular reviewer for several international journals. He is a Senior IEEE member.

His research interests include parametrized modeling and model order reduction, dynamical systems, machine learning, sampling techniques, design space exploration, uncertainty quantification, optimization, and behavioral modeling.
\end{IEEEbiographynophoto}

\vfill


\end{document}